\documentclass[showpacs,preprint]{revtex4}
\usepackage{graphics}
\usepackage{epsf}
\usepackage{amsfonts}
\usepackage{amssymb}

\catcode `@=11 \@addtoreset{equation}{section} \catcode `@=12

\def\tsector#1#2{\ {\scriptstyle #1}\hskip 1mm
\mathop{\opensquare}\limits_{\lower 1mm\hbox{$\scriptstyle#2$}}^\sim\hskip 1mm}
\def\appendix{{\newpage\section*{Appendix}}\let\appendix\section%
        {\setcounter{section}{0}
        \gdef\thesection{\Alph{section}}}\section}
\def\){\right)}
\def\({\left( }
\def\]{\right] }
\def\[{\left[ }

\def\half{{1\over 2}}

\newcommand{\be}{\begin{equation}}
\newcommand{\ee}{\end{equation}}
\newcommand{\ba}{\begin{eqnarray}}
\newcommand{\ea}{\end{eqnarray}}
\newcommand{\no}{\nonumber \\}

\def\Tr{{\rm Tr}}
\def\tr{{\rm tr}}

\newcommand{\cN}{{\cal N}}

\begin{document}
\title {{\small \hfill SLAC-PUB-10180 }~~~~~~~\\[3ex]
Matrix model, Kutasov duality and Factorization of Seiberg-Witten Curves }
\author{ Matthias Klein$^a$ } \author{Sang-Jin Sin$^b$ } 
  
\affiliation{\small $^a$\,Stanford Linear Accelerator Center, Stanford University, Stanford CA 94309 }
\affiliation{\small $^b$\,Department of Physics, Hanyang University, Seoul, 133-791, Korea  } 

\begin{abstract}  
We study the duality of $\cN=1$ gauge theories in the presence of a massless adjoint field and massive quarks by calculating the superpotential using the  Dighkgraaf-Vafa matrix model  and by comparing with the previous result coming from Kutasov duality. 
The Kutasov duality method gives a result in which one instanton term is absent. The matrix model method confirms it and also show that the absence of the one instanton term is related to the masslessness of the adjoint field. 
\end{abstract}
\maketitle
\newpage
\section{Introduction}

After ground breaking work on $\cN=2$ \cite{sw} and $\cN=1$ supersymmetric 
(SUSY) gauge theories \cite{seiberg}, duality became the central issue in 
theoretical physics and thousands of works have appeared. One of the most 
natural extension of the work is to include adjoint fields. In fact Kutasov 
and his collaborators extended the notion of Seiberg duality \cite{seiberg} 
to such cases \cite{kutasov} and a brane picture \cite{kutasovreview} of this
duality, which we call Kutasov duality for brevity, has been worked out.   
However, compared with $\cN=2$ gauge theory, not much quantitative information 
had been worked out for $\cN=1$ theory until recently. 

Recently, Dijkgraaf and Vafa considered $\cN=1$ theory as a deformation of 
$\cN=2$ theory by a superpotential for the adjoint field and discovered a 
surprising link between effective superpotentials of SUSY gauge theories 
and those of the associated bosonic matrix model \cite{dv}.
This link allows one to obtain 
nonperturbative results by doing a perturbative calculation.
It is a powerful method to compute the effective superpotential
of ${\cal N}=1$ gauge theories with massive matter in tensor and 
fundamental representations. However, it is difficult to use
this method if the gauge theory contains massless adjoint fields.\footnote{See 
however \cite{Feng}, where the case of massless {\it fundamental} fields is 
discussed.}
More recently, it is understood \cite{cdsw} (see also \cite{dvg}) that such 
a simplification is due to the super symmetries by considering (anomalous) 
Ward identities in gauge theory  is the same as the loop equation in the 
matrix model.  Soon after, it is  shown \cite{csw1,csw2,nsw1,nsw2} that the 
the loop equation is identical to the factorization of the Seiberg-Witten 
curve of the original ${\cal N}=2$ theory to the reduced curve defined by the 
superpotential.\footnote{More precisely, the loop equation is equivalent to 
the minimization of the superpotential, which requires that all periods of the 
generating 1-form  on the Riemann surface defined by the superpotential are  
integers. 
By Abel's theorem, the 1-form must be a derivative of a meromorphic function 
$\psi$. Finally, the condition that $\psi$ be single valued on the reduced 
Riemann surface is equivalent to the factorization condition.} This completes 
the recipe for the solution to the problem. It corresponds to an extension of 
the earlier results of Vafa and his collaborators \cite{civ,cv} on the problem 
without fundamentals. 

In a recent paper \cite{KS}, we utilized Kutasov duality and the 
gauge theoretic method \cite{seiberg,si,murayama} to find the superpotential 
of $SU(N_c)$ gauge theory with $N_f$ massive fundamental fields and a massless 
adjoint field having nontrivial tree level superpotentials. More specifically, 
for a theory with tree level superpotential
\be  \label{Wtree}
W_{tree}=\sum_{l=1}^{2}\Tr(m_l\,\tilde Q \Phi^{l-1} Q)
       + {\textstyle \frac{1}{3}}\,g\, 
         \tr\Phi^{3},
\ee
we found that for the unbroken gauge symmetry, the effective superpotential is 
\be
W_{eff}=g\,\Lambda_L^4\,
             \Tr\left(m_2 m_1^{\ -1}\right),
\ee
To get this result, we first worked out the case where both the adjoint as well as the fundamental fields are massless. Then Kutasov duality was used 
to map the result to the case where the fundamental fields are massive.

One interesting fact to notice is that there is no term proportional to $\Lambda_L^2$, which is a one instanton contribution,  while it is expected from the naive 
dimensional analysis.  
The purpose of this paper is to calculate the superpotential
by Dighkgraaf-Vafa Matrix model technology 
\cite{civ,cv,csw1,janik,bala} and confirm that this is indeed so.  We will also show that the absence of the one instanton term is related to the masslessness of the adjoint field. 

The rest of the paper goes as follows: in section 2, we will briefly review 
the factorization method used in this paper. In section 3, we calculate 
the superpotential with cubic tree level potential in the presence of the fields in fundamental representation and show the absence of the one instanton contribution. 
Section 4 gives the conclusion. 
In the appendix, we give the superpotential for two cut case for later interests and also an account for the equivalence of factorization of the Seiberg-Witten curve and the minimization of the superpotential by an explicit calculation for simple cases.

\section{Brief review: matrix model and factorization}
Here we give a review of the necessary material.
We start from the matrix model partition function 
\be 
Z=\frac{1}{vol(G)}\int d\Phi dQ^id{\tilde Q}^i \exp\(-\frac{1}{g_s}W (\Phi)- 
\frac{1}{g_s}\sum_{i=1}^{N_f}\[{\tilde Q_i} \Phi Q^i -m_i {\tilde Q_i}Q^i\]\),
\ee
where $W (z)$ is a polynomial of order $n+1$.
Diagonalizing $\Phi$ and integrating over $Q$ and $\tilde Q$, we get 
\be
Z \sim \int \prod_{a=1}^N \exp\(-\frac{1}{g_s}W (\lambda_a) 
+2\sum_{a<b}^N \log(\lambda_a-\lambda_b)
-\sum_{i=1}^{N_f}\sum_a\log(\lambda_a-m_i)\)
\ee
 The saddle point equation 
in the  limit $g_s\to 0$ and $N\to \infty$ with $S=g_sN$ fixed, 
 is called the loop equation:
\be
S\omega^2(z)+ {W'_0(z)} \omega(z)+\frac{1}{4}f(z)=0,\ee
where $\omega(z)$ is the resolvant defined as  
\be
\omega(z)=\frac{1}{N} \tr \( \frac{1}{\Phi-z} \)=
\frac{1}{N}\sum_a\frac{1}{\lambda_a-z}
\ee
and $f$ 
is a polynomial of order $n-1$ yet to be determined.
  
The large $N$ can be expressed in terms of density of eigenvalues  
$\rho(\lambda)=\frac{1}{N}\sum_a\delta(\lambda-\lambda_a)$  
 normalized by $\int\rho(\lambda)=1$. In terms of it,  
$\omega(z)=\int d\lambda\frac{\rho(\lambda)}{\lambda-z}$ and 
\be
\rho(\lambda)=\frac{1}{2\pi i}[\omega(\lambda+i\epsilon)
              -\omega(\lambda-i\epsilon)]
= \frac{1}{2\pi i} {\rm Disc}{[\omega(z)]}.\label{disc}\ee

The loop equation can be rewritten as a defining equation of a 
hyperelliptic Riemann surface 
\be 
y^2=W'(z)^2+f(z), \quad{\rm with}  \quad y(z)=2S\omega(z)+W '(z).
\ee
The curve has $n$ cuts in the $z$ plane. 
The eigenvalues of $\Phi$ are distributed along the cuts according to $f$.  
Let $N_i$ be the number of eigenvalues along the $i$-th cut: 
$N_i=N\int d\lambda\rho(\lambda)$, and let $S_i:=g_sN_i$ be finite  
in the limit $N\to\infty,$ $g_s\to 0$. Using the eq.(\ref{disc}), 
the latter can be rewritten as
$S_i=-\frac{1}{4\pi i}\oint_{A_i} y(z) dz$, where $A_i$ is a contour 
encircling the $i$-th cut.
Following \cite{cv,nsw2}, we denote by $P,Q$ the point $z=\infty$ on the 
two sheets of the hyperelliptic curve such that $y(P)\sim W'_0(P)$. 
The hyperelliptic curve can be given canonical homology cycles 
$A_i (i=1,...,n-1)$ and $B_i={\hat B}_i-{\hat B}_n (i=1,...,n-1)$. 
When $n=N_c$, $f$ and the Seiberg-Witten curve were determined in 
\cite{nsw2} following the work in  \cite{cv}.
For $n<N_c$, this problem is solved in \cite{nsw2,csw2}.
The result is   
\be 
W_{eff}=\sum_{l=1}^n N_{c,i}\frac{\partial F_s}{\partial S_i} 
        +F_d +2\pi i\tau_0 \sum_{l=1}^{n}S_i.\ee
One can easily express the relevant quantities in terms of the integral 
of $y$'s over the infinite cycles,
\be
\frac{\partial F_s}{\partial S_i} =-\half \int_{{\hat B}_i}ydz
                                  = -\int_{e_i}^P ydz ,\quad 
F_d= \half \int_{m_i}^Py dz,
\ee
up to the integral constants cancelling the divergences, which are 
independent of $S$. Here, the $e_i$'s are the boundaries of the cuts. 
Finally we get 
\be
W_{eff}=-\sum_{l=1}^n N_{c,i}\(\int_{e_i}^P ydz-W(P)\) 
        + \half \(\int_{m_i}^Py dz-W(P)+W(m_i)\)
        +2\pi i\tau_0 \sum_{l=1}^{n}S_i.\ee
What is shown in \cite{nsw2,csw2} is that the minimization of this 
superpotential is equivalent to the integer periodicity 
along $A_i$ and $B_i$ cycles. By Abel's theorem, 
the 1-form $\omega(z)$ must be a derivative of a meromorphic 
function $\psi$, i.e., $\omega (z)dz=d\psi$. For $N_f<2N_c$, 
$\psi=P(z)+\sqrt{P^2(z)-\alpha B(z)}$. 
Finally, the condition that $\psi$ be single valued on the reduced Riemann 
surface $y^2=W'^2+f$ is the factorization of the Seiberg-Witten 
curve of the original $\cN=2$ theory to the reduced curve defined by the 
superpotential. 
\ba
P_N^2(z)-\alpha B(z)&=& F_{2m}(z)H_{N-m}^2(z),\no
W_{n+1}'^2(z)+f_{n-1}(z) &=& F_{2m}(z)Q^2(z).\label{factor}
\ea
The sub-indices of the polynomials are their orders and 
the number of cuts $m$ in eigenvalue space is related to the order of 
$F_{2m}$. 
This completes the recipe for the solution to the problem and 
corresponds to  extending the earlier result of Vafa and his 
collaborators \cite{civ,cv} on the problem without fundamentals. 

\section{Absence of one instanton term via Dighkgraaf-Vafa}
We now evaluate the superpotential using the factorization method. 
We  consider the simplest nontrivial case: $N_c=3$ and $N_f=2$. 
We take the superpotential of the adjoint field $\Phi$ to be 
\be
W =\frac{g}{3}\tr\Phi^3+\frac{m_\phi}{2}\tr\Phi^2+{\lambda}\tr\Phi,
\ee
such that  
 \be
 W'(z)=gz^2+m_\phi z+\lambda.\ee
Since $W'$  is of degree two, $y=\sqrt{W'^2+f_1}$ will have at most 
two cuts and the relevant Riemann surface is of genus one, allowing an
explicit study in terms of well known technology on the torus.
 
 Since the gauge theory result is for unbroken gauge group, we 
consider the one-cut solution. The condition is 
\be W'^2_3+f=F_2\cdot Q_1^2.\ee
   It is easy to see that number of unknowns 
is bigger than number of equations by 1. For a cubic potential, 
$Q$ must be linear. Hence we put $Q=x-x_3$. Then 
\be
  (x^{2} + {\displaystyle \frac {{m_{\phi }}\,x}{g}}  + 
{\displaystyle \frac {\lambda }{g}} )^{2} + {\displaystyle 
\frac {{f_{1}}\,x + {f_{0}}}{g^{2}}} =(x - {x_{1}})\,(x - {x_{2}})\,(x - {x_{3}
})^{2} 
\ee
We introduce the variables $\Delta:=(x_1-x_2)/2$ and $T:= (x_1+x_2)/2$. 
After some algebra, we can show that $x_3$ can be determined by $\Delta$ 
from the equation
\be
W'(x_3)+\half g\Delta^2=0,
\ee
and $T,f_0,f_1$ can be determined in terms of $\Delta, x_3$ as follows;
\be
T= -  m_{\phi }/g + x_3,\; {f_{1}}=(g\Delta)^2(2x_3+{m_\phi\over g}),\; 
f_0= g\Delta^2(m_\phi x_3+2\lambda+\frac{3}{4}g\Delta^2).
\ee
To determine $\Delta$, we must use the  factorization condition,
\be
P^2(x)-4\Lambda^{2N_c-N_f}\prod_{i=1}^{N_f}(x-m_i)=F_2(x)H^2(x).\ee
For  $N_c=3, N_f=2$, and setting $m_1=m_2=m$, we have the factorized version
\ba
P(x)-2\,\epsilon
 \,\Lambda ^{2}\,(x - m)&=&(x - {x_{1}})\,(x - {h_{1}})^{2},\no
P(x)+2\,\epsilon
 \,\Lambda ^{2}\,(x - m)&=&(x - {x_{2}})\,(x - {h_{2}})^{2}.\label{factor32}
 \ea
 Since we are mainly interested in the $SU(3)$ case, we impose the 
traceless condition explicitly:
\be
x_1+2h_1=0, \quad x_2+2h_2=0.
\ee
Then eq. (\ref{factor32}) gives 
\be
3T\Delta={4}\epsilon\Lambda^2, \;\;\;\;
{3}T^2 \Delta+ \Delta^3+8\epsilon\Lambda^2 m=0,\ee
and $\Delta$ is determined by
\be
\Delta^4+8\epsilon m\Lambda^2\Delta+\frac{16}{3}\Lambda^4=0.\ee
Similarly, $T$ is determined by 
\be 
T^4+2mT^3+\frac{16}{27}\Lambda^4=0.
\ee
For $\Lambda\ll m$, we can solve for $\Delta$ and $T$ 
in terms of a power series in $\Lambda$. There are two real solutions:
One solution is of integer powers in $\Lambda$; 
\ba
\Delta&=&-\frac{2}{3}\frac{\Lambda^2}{m}-\frac{2\Lambda^6}{81m^5}
         -\frac{8\Lambda^{10}}{2187m^9}-\frac{44\Lambda^{14}}{59049m^{13}}
         +O(\Lambda^{18}), \no
T&=&-2m+\frac{2\Lambda^4}{27m^3}+\frac{2\Lambda^8}{243m^7}+
\frac{10\Lambda^{12}}{6561m^{11}}+O(\Lambda^{16}). \label{sol1}
\ea
It turns out that this solution  does not correspond to the 
result obtained in \cite{KS} by gauge theory.   The reason is 
is simple: In gauge theory, we usually consider the unbroken gauge group, so that the roots of $P_3$ must be completely degenerate in $\Lambda \to 0$ limit. Therefore we should look for $T$ which vanishes in that limit. 
The other (correct) solution is of fractional power in $\Lambda$:
\be
T\sim -(2/3)(\Lambda^4/m)^{1/3} \quad {\rm and} \;\; 
\Delta\sim -2(\epsilon m\Lambda^2)^{1/3}.\ee 
In order to understand this  solution, we introduce
${ \Lambda_L}^3=m\Lambda^2$.   
In fact $\Lambda_L$ is precisely the low energy quantum scale  
appearing in the  field theory analysis that is defined by
\be 
\Lambda^{2N_c-N_f}\det m={\Lambda}_L^{2N_c}.
\ee 
In terms of $\Lambda_L$, the second  solution is analytic in $\Lambda_L$ and given by
\ba
T=- {\displaystyle \frac {2\,\Lambda_L^{2}}{3\,m}
}  - {\displaystyle \frac {2\,\Lambda_L ^{4}}{27\,m^{3}}}  - 
{\displaystyle \frac {2\,\Lambda_L ^{6}}{81\,m^{5}}}  &-&  
{\displaystyle \frac {70\,\Lambda_L ^{8}}{6561\,m^{7}}}  - 
{\displaystyle \frac {308\,\Lambda_L ^{10}}{59049\,m^{9}}}  - 
{\displaystyle \frac {2\,\Lambda_L ^{12}}{729\,m^{11}}}    \no
&-& {\displaystyle \frac {21736\,\Lambda_L ^{14}}{14348907\,m^{13}}}
- {\displaystyle \frac {111826\,\Lambda_L ^{16}}{129140163
\,m^{15}}}+O(\Lambda_L^{18}), \label{TL}\ea
\ba
\epsilon \Delta=- 2\,\Lambda_L   &+& {\displaystyle \frac {2}{9\,m^{
2}}} \,\Lambda_L ^{3}+{\displaystyle \frac {4}{81\,m^{4}}} \,
\Lambda_L^{5} 
+ {\displaystyle \frac {40}{2187\,m^{6}}} \,\Lambda_L 
^{7}+ {\displaystyle \frac {2}{243\,m^{8}}} \,\Lambda_L ^{9} \no 
&+&
{\displaystyle \frac {728}{177147\,m^{10}}} \,\Lambda_L ^{11} + 
{\displaystyle \frac {10472}{4782969\,m^{12}}} \,\Lambda_L ^{13} 
+ {\displaystyle \frac {8}{6561\,m^{14}}} \,\Lambda_L^{15} + 
O(\Lambda_L^{17}). \label{DL}
\ea

Now, we can calculate the superpotentials. 
 For the massless adjoint field, which is the case we consider,  the superpotential is given by  
\be
W_{eff}=gu_3=\frac{g}{4}(T+\Delta)^3-2g\epsilon\Lambda^3_L, \label{weff}
\ee
as a power series in $\Lambda_L$:
\ba
W_{eff}= - {\displaystyle \frac {2\,g}{m}} \,{\Lambda_{L}}^{4} 
         + {\displaystyle \frac {4\,g}{27\,m^{3}}} \,{\Lambda _{L}}^{6} 
   &+&   {\displaystyle \frac {2\,g}{81\,m^{5}}} \,{\Lambda _{L}}^{8} 
    + {\displaystyle \frac {16\,g}{2187\,m^{7}}} \,{\Lambda _{L}}^{10} \no 
   &+&   {\displaystyle \frac {2\,g}{729\,m^{9}}} \,{\Lambda _{L}}^{12} 
    + {\displaystyle \frac {208\,g}{177147\,m^{11}}} \,{\Lambda _{L}}^{14} 
    + \mathrm{O}({\Lambda _{L}}^{16}).
\ea
Notice the absence of a term proportional to $\Lambda_L^2$.
The superpotential indeed start from the two instanton contribution ($\Lambda_L^4$ as we aimed to prove. 
The result obtained by Kutasov duality \cite{KS} 
is just the leading term of what is found here. 
Since the duality is just IR equivalence of two theory,
the Kutasov duality can not say about the  higher order. \footnote{
Notice also that  there is no term with an odd power of $\Lambda_L$. 
Considering the structure of eq.\ (\ref{weff}) and the $\Lambda$ dependence of 
$T$, $\Delta$ in eq.\ (\ref{TL}),(\ref{DL}), this result seems to be rather
 nontrivial. It is not very clear to us what symmetry causes such result. }

One should mention that the absence of the one instanton term  is  a consequence of the masslessness of the adjoint field. 
In order to show this, we consider the  the case  where the adjoint field is massive. For simplicity we only consider in the $SU(3)$ theory only. Then, 
$W_{eff}=gu_3+m_\phi u_2$.
Using 
\be
u_2:= 3(T+\Delta)^2/4-2\epsilon\Lambda_L^3/m,
\ee
the superpotential is calculated to be
\ba
W_{eff}=&&3\,{m_{\phi }}\,{\Lambda _{L}}^{2} + ( - 
{\displaystyle \frac {2\,g}{m}}  - {\displaystyle \frac {1}{3}} 
\,{\displaystyle \frac {{m_{\phi }}}{m^{2}}} )\,{\Lambda _{L}}^{4
} + ({\displaystyle \frac {4\,g}{27\,m^{3}}}  - {\displaystyle 
\frac {1}{27}} \,{\displaystyle \frac {{m_{\phi }}}{m^{4}}} )\,{
\Lambda _{L}}^{6} \no 
&+& ({\displaystyle \frac {2\,g}{81\,m^{5}}}  - 
{\displaystyle \frac {7}{729}} \,{\displaystyle \frac {{m_{\phi }
}}{m^{6}}} )\,{\Lambda _{L}}^{8}  
 + ({\displaystyle \frac {16\,g}{2187\,m^{7}}}  - {\displaystyle 
\frac {22}{6561}} \,{\displaystyle \frac {{m_{\phi }}}{m^{8}}} )
\,{\Lambda _{L}}^{10} + ({\displaystyle \frac {2\,g}{729\,m^{9}}
} \no
 &-& {\displaystyle \frac {1}{729}} \,{\displaystyle \frac {{m_{
\phi }}}{m^{10}}} )\,{\Lambda _{L}}^{12} +  
({\displaystyle \frac {208\,g}{177147\,m^{11}}}  - 
{\displaystyle \frac {988}{1594323}} \,{\displaystyle \frac {{m_{
\phi }}}{m^{12}}} )\,{\Lambda _{L}}^{14} + \mathrm{O}({\Lambda _{
L}}^{16}).
\ea
Note that $\Lambda_L^2$ is non-vanishing unless the adjoint field is massless, as we claimed.

\section{Conclusion}
In this paper, we   calculated the superpotential for an 
$SU(3)$ gauge theory in the presence of  massive fundamental fields and a massless adjoint field having nontrivial tree-level superpotentials  using the method of factorizing the Seiberg-Witten curve, and  verified the absence of the one instanton contribution, which was suggested by the Kutasov duality. We also  showed that the absence of the one instanton term is related to the masslessness of the adjoint field. 
  It would be very interesting to understand the these connections in more fundamental way. 
 We wish to come back to these issues later. 
 
 \begin{acknowledgments} 
We would like to thank C.~Cs\'aki, H.~Murayama and R.~Tatar for helpful discussions.
The work of SJS is  supported by KOSEF Grant 1999-2-112-003-5. 
The work of MK is supported by the Deutsche Forschungsgemeinschaft.
\end{acknowledgments}

\appendix {Two-cut solution }
Consider an $\cN=1$ theory with superpotential $W$ as a deformation 
of an $\cN=2$ theory with particular moduli parameters for which the 
following equation holds:
\be P_{N_c}(x;u_p)^2-4\Lambda^{2N_c-N_f}\prod_{i=1}^{N_f}(x-m_i)
    =H_{N-n}^2(x)\frac{1}{g^2_{n+1}}(W'^2+f), \ee
where $P_N(x)=\det(x-\Phi)$ for $N_f<N_c$.  \footnote{For $N_f>N_c$, the  factorization equation should be replaced by:
$$ (P_N(x;u_p)+{1\over 4}\Lambda^{2N_c-N_f}Q_{N_f-N_c})^2
    -4\Lambda^{2N_c-N_f}\prod_{i=1}^{N_f}(x-m_i)
    =H_{N-n}^2(x)\frac{1}{g^2_{n+1}}(W'^2+f),
$$ with 
$Q=\sum_i^{N_f-N_c}x^{N_f-N_c-i}t_i(m)$. }
This corresponds to the $Q=1$ case in (\ref{factor}).
This equation dictates that, apart from the $n$ photons, there should be 
$N-n$ extra massless fields  like monopoles and dyons.
Notice that if we let the coefficient of highest power be 1, there are 
$N_c+(N_c-n)+n$ parameters and $2N_c$ equations. Therefore, given $W'$ and 
the matter part $4\Lambda^{2N_c-N_f}\prod_{i=1}^{N_f}(x-m_i)$, both the 
moduli of the Seiberg-Witten curve, $P_N$, as well as those of the monopoles,  
$H_{N-n}$, are uniquely fixed. 
For the parameters satisfying this factorization, the glueball fields 
$S_i$ are also determined in terms of the classical data of $W , m_i$ 
and the quantum scale $\Lambda$.  

 For our case $N_f=2, N_c=3$, we carry out the factorization under the 
assumption that all fundamental hypermultiplets have the same mass $m$ as 
before: 
\be
P_3^2-4\Lambda^4(x-m)^2=(x-a_1)^2(W'^2+f)/g^2.
\ee
Notice that one of $P_3\pm 2\Lambda^2(x-m)$ must have the factor $(x-a_1)^2$.
 Let 
\be
P_3-2\epsilon\Lambda^2(x-m)=(x-a_1)^2(x-a_2),\ee
with $\epsilon=\pm 1$. 
Then,
\be
P_3+2\epsilon\Lambda^2(x-m)=(x-a_1)^2(x-a_2)+4\epsilon\Lambda^2(x-m).\ee
Therefore we can identify 
\be
(W'^2+f)/g^2=(x-a_1)^2(x-a_2)^2+4\epsilon\Lambda^2(x-m)(x-a_2).
\label{ysqare}\ee
Since $f$ is at most linear in $x$, we can determine $f$ and $a_i$ in terms 
of classical data and $\Lambda$ by the  identification
\ba
W'/g&=&(x-a_1)(x-a_2)+2\epsilon\Lambda^2,\label{w'} \cr
f(x)/g^2&=&4\epsilon\Lambda^2(a_1-m)(x-a_2)-4\Lambda^4.\label{f}
\ea
Therefore $a_1,a_2$ are solutions of 
\be 
W'/g-2\epsilon\Lambda^2=0, \label{eqm}
\ee

\be
a_1,a_2=-\frac{m_\phi}{2g} \pm \sqrt{\frac{m_\phi^2}{4g^2}-\frac{\lambda}{g} 
+2\epsilon\Lambda^2} \label{a1a2}.\ee
Since $f=-4g Sx+f_0$, the factorization determines the exact value of gluino
condensate:   
\be
S=\epsilon g \Lambda^2 \(-\frac{m_\phi}{2g} -m+\sqrt{\frac{m_\phi^2}{4g^2}
-\frac{\lambda}{g} +2\epsilon\Lambda^2}\).\ee
To calculate the superpotential in this case, 
we use the method of \cite{civ}.
\be
P_3(x)=(x-a_1)^2(x-a_2)+2\epsilon\Lambda^2(x-m)
      :=  \sum_{i=0}^{N_c=3} s_{N_c-i}x^i.\ee
Using Newton's relation, $ks_k+\sum_r ru_rs_{k-r}=0$,  
\be
u_1 =-s_1, \;\; u_2 =-s_2 +\half s_1^2,\,\;u_3 = -s_3+s_2s_1-{1 \over 3}s_1^3.
\ee
We get 
\ba
u_1&=&2a_1+a_2 ,\no
u_2&=&\half(2a_1^2+ a_2^2)-2\epsilon \Lambda^2 , \no
u_3&=&{1\over3}(2a_1^3+ a_2^3)-2\epsilon \Lambda^2(2a_1+a_2-m) .
\ea
Therefore\footnote{Notice that the minimization of $W_{eff}$ w.r.t.\ $a_i$ 
gives $W'(a_i)=2g\epsilon \Lambda^2, \;\; {\rm for} \;\;  i=1,2,$
which is precisely  equal to the factorization result (\ref{eqm}). 
This method was first used in \cite{civ} and can be used in the present
case with fundamentals without much change.}
\ba
W_{eff}&=&gu_3+m_\phi u_2+\lambda u_1 \no &=&2W(a_1)+W(a_2)
              -2\epsilon \Lambda^2(m_\phi+g(2a_1+a_2-m)).
\ea
After some calculation using (\ref{a1a2}), we get 
\be
W_{eff}=-\frac{3}{2}\frac{\lambda m_\phi}{g}+ \frac{1}{4}\frac{m^3}{g^2}
        + (m_\phi+2gm)\epsilon\Lambda^2 
        +\(\frac{2}{3}\lambda-\frac{m_\phi^2}{6g} 
        -\frac{4}{3}g\epsilon\Lambda^2\) \sqrt{\frac{m^2_\phi}{4g^2}
        -\frac{\lambda}{g}+2\epsilon\Lambda^2}
\ee 
This factorization result contains both the disk as well as the sphere contribution as argued in \cite{janik}.  
  
In the case of $SU(3)$,  we have to impose the traceless condition:
$u_1=dW_{eff}/d\lambda=2a_1+a_2=0$. Together with $a_1+a_2=-m_\phi/g$, this gives us much simpler results;
\be
 a_1=m_\phi/g, \quad a_2=-2m_\phi/g, \quad  \lambda=2g\epsilon\Lambda^2-2m^2_\phi/g.
\ee
Notice that $a_1,a_2$ became the same as the classical solution of $W'$.
Then, the gluino condensate is given by
\be
S=\epsilon \Lambda^2\({m_\phi}-mg\), \;{\rm for}\;\; SU(3).
\ee 
The superpotential for $SU(3)$ can be simplified to
\be
W_{eff}=\frac{m^3_\phi}{g^2}-2\epsilon\Lambda^2(m_\phi-gm).
\ee

\section{Factorization vs. minimization}
For later interests, we give an account for the equivalence of the two methods by explicit computations for simple cases.  We concentrate mostly $U(2)$, $SU(2)$ SYM with fundamentals and  quadratic superpotential 
for the adjoint field $\Phi$:
\be
W=\half m_\phi \Phi^2 +\lambda  \Phi,
\ee
so that 
\be
W'(x)=m_\phi x+\lambda, \; f=-4Sm_\phi, \label{f0}
\ee where $S$ is yet to be determined. 

{$\bullet$ Factorization for $N_c=2$ and $N_f=1$}

Let $x_1,x_2$ be the solutions of $W'^2+f=0$ and $\alpha, \beta$ be the 
those of $P_2(x)=0$. 
Then with the identification 
\be 
T:=(x_1+x_2)/2=-\lambda/m_\phi \;\; {\rm and } \;\; \
\Delta=:(x_1-x_2)/2=\sqrt{2S/m_{\phi}},\ee
 we can set  
\be (x - \alpha )^{2}\,(x - \beta )^{2} -4\,
\Lambda ^{3}\,(x - m) = \((x -T)^{2} -\Delta^2 \)
\,(x - a)^{2}
\ee
This identity leads us to 
\ba
\alpha  + \beta  &=& T + a,  \no
 (\alpha+\beta)^2 +2 \alpha  \beta &=& T^{2} - \Delta ^{2} + 4 T a + a^{2},\no
 \alpha \beta(\alpha +\beta) &=& -2 \Lambda ^{3} 
+  (T^{2} - \Delta ^{2}) a +  T a^{2} , \no
   \alpha ^{2} \beta ^{2} &=& - 4 \Lambda ^{3} m + (T^{2} - \Delta^{2}) a^{2} .
\ea
For the $U(2)$ case, $\lambda$ is known, so is $T$. 
Eliminating $\alpha$ and $\beta$, we get 
\be
(T-a)\Delta^2=4\Lambda^3, \;\; 
(\Delta/2)^6+(m-T)(\Delta/2)^2+ \Lambda^6=0. \label{delta}
\ee
Combining these two, we can also get 
\be 
a^2-(T+m)a+mT+\frac{1}{4}\Delta^2=0,
\ee
so that 
\be
a=\half\(m+T\pm\sqrt{(m-T)^2-\Delta^2}\),
\ee
One needs to choose the $-$ sign from the  $T=0$ behavior 
$a\sim \Delta^2/(4m)$ in the limit $\Lambda\to 0$. 
Then, 
\be
\Lambda^3=\frac{\Delta^4/8}{-m+T- \sqrt{(T-m)^2-\Delta^2}}.
\ee 
Let's calculate the superpotential $W_{eff}=m_\phi u_2+\lambda u_1$. 
From $u_1=-s_1=\alpha+\beta=a+T$ and 
$u_2=-s_2+\half s_1^2=-\alpha\beta +\half u_1^2=\half(a^2+T^2+\Delta^2)$, 
\be
W_{eff}= m_\phi\( \frac{1}{2}(m-T)a-\half T^2+\frac{3}{8}\Delta^2\)
.\ee
Since the value of $\Delta^2$ can be obtained in terms of $\Lambda, T, m$ 
from the second equation of eq. (\ref{delta}), this completes the solution. 
To compare with the known result, we set $\lambda=T=0$. Then 
in terms of  $S$ and  $\alpha=1/(m_\phi m^2)$,
\be 
W_{eff}=\frac{3}{2}S+\frac{1-\sqrt{1-4\alpha S}}{4\alpha},
\ee
 which precisely agrees with the known super potential \cite{bena}
 at its critical value. 

For the $SU(2)$ case, $\lambda$ is unknown but the traceless condition  
$\alpha+\beta=0$ determines it.
In this case, $a$ and $\Delta$ are determined by 
\be
a^2-ma+\frac{1}{8}\Delta^2=0, \quad a\Delta^2=-2\Lambda^3.
\ee
In terms of $\Lambda_L$ defined by $\Lambda_L^4=m\Lambda^3$, 
the combined equation can be written as 
\be
a^3-ma^2-\frac{1}{4m}\Lambda_L^4=0,\ee
which gives $a$ in series  of $\Lambda_L$:
\be
a=\half i\Lambda_L^2/m-\frac{1}{8m^3}\Lambda_L^4-\frac{5i}{64m^5}\Lambda_L^6+\frac{1}{16}m^7
\Lambda_L^8+\frac{231i}{4096m^9}\Lambda_L^{10}-\frac{7}{128m^{11}}\Lambda_L^{12}+ O(\lambda_L^{14}) \ee
Then $T=-a$, $\Delta^2=-2\Lambda_L^4/(ma)$ and $S=\frac{1}{4}m_\phi\Delta^2$ give the desired results.
Here again it is interesting to observe that the series is of even powers of $\Lambda_L$.

{$\bullet$ $N_c=2$ and $N_f=2$}

We  consider the case where the two $m_i$ are equal to $m$.
Using the Seiberg-Witten curve for $N_c=N_f$ given in   \cite{hananyoz}, the  
factorization condition is 
\be 
y^2=P^2 - 4\Lambda^2(x-m)^2=(x-a)^2F_2,
\ee
where $P=P_2+\delta\Lambda^2$ and $\delta$ is usually $1/4$ but $1/8$ for 
$N_c=2$. 
Therefore
\be
P-2\epsilon\Lambda(x-m)=(x-a)^2.
\ee
Then 
\ba
F_2&=&(x-a)^2+4\epsilon\Lambda(x-m) \no
    &=&(W'^2+f)/m^2_\phi=(x+\frac{\lambda}{m_\phi})^2-{4S\over m_\phi},
    \ea 
which leads us to the relations
$4\epsilon \Lambda-2a=2\lambda/m_\phi$, and $a^2-4m\epsilon\Lambda
=\frac{\lambda^2}{m^2_\phi} -\frac{4S}{m_\phi}$.
From this we get $a=2\epsilon\Lambda-\frac{\lambda}{m_\phi},$ and more 
importantly for our purpose, we determine the gaugino condensate  
\be
 \quad S=\epsilon\lambda\Lambda-m_\phi\Lambda^2+\epsilon mm_\phi\Lambda. 
\label{S}
\ee
Now let's move to determine the superpotential.
\be
P_2=(x+\frac{\lambda}{m_\phi})^2-\frac{4S}{m_\phi}-2\epsilon\Lambda(x-m)
-\delta\Lambda^2.
\ee
Hence
\ba
u_1&=&2(\epsilon\Lambda-\frac{\lambda}{m_\phi}),\no
u_2&=& \(\frac{\lambda}{m_\phi}\)^2-2\Lambda^2+2\epsilon m\Lambda 
      -\delta \Lambda^2.
\ea
Now,
\ba
W_{eff}&=&m_\phi u_2+\lambda u_1 \no
            &=& -\frac{\lambda^2}{m_\phi}+ 
        2\epsilon(\lambda\Lambda+mm_\phi\Lambda)-(2+\delta )m_\phi\Lambda^2.
\ea

For comparison with previous literature 
we also consider 
$U(2)$ with $\lambda=0$;
\be
a=2\epsilon\Lambda, \quad S= -m_\phi\Lambda^2+\epsilon mm_\phi\Lambda,
\ee
and
\be 
W_{eff}=2\epsilon mm_\phi\Lambda-(2+\delta )m_\phi\Lambda^2
=2S-\delta m_\phi\Lambda^2.
\ee

For $SU(2)$, by imposing $u_1=0$,  $\lambda$ is determined  and $a$ and $S$ 
are simplified  so that we have 
\be
\lambda=\epsilon m_\phi\Lambda,\quad a=\epsilon \Lambda, \;\; S
=\epsilon mm_\phi\Lambda.
\ee
The  effective potential is 
\be
W_{eff}=2\epsilon mm_\phi\Lambda-(1+\delta )m_\phi\Lambda^2
.\ee
 
{$\bullet$ \it Minimization }

We apply the loop equation to the case where the exact result is known.
$W_0=\half m_{\phi} \tr \Phi^2$.
So the tree-level superpotential is 
\be
W=W_0+\sum_{i=1}^{N_f}({\tilde Q}_i\Phi Q_i + {\tilde Q}_i m Q_i).
\ee 

Then the surface equation is given by 
\be
y^2=W^{'2}+f= (m_{\phi}z)^2 - 4Sm_{\phi}.
\ee
So, 
\be
W^{(s)}_{eff}=-N_c\(\int_e^P y dz -W(P)\) 
           = N_cS(1-\log {S\over m_{\phi}\Lambda_0^2}),\ee
 where $ e^2=4S/m_\phi$ and $\Lambda_0$ is the location of $P$.  
Similarly,
\ba
 W^{(d)}_{eff} &=&  \half\sum_{i=1}^{N_f}\(\int_{m_i}^P y dz -W(P)+W(m_i)\)\no
 &=& {N_f}S\log \(\frac{m}{\Lambda_0} \)  
     -N_fS \(\half+\frac{\sqrt{1-4\alpha S}-1}{4\alpha S} 
     -\log\frac{1+\sqrt{1-4\alpha S}}{2} \),
 \ea
 where $\alpha=1/(m^2m_{\phi})$.

By minimization condition of the total superpotential  gives
\be 
\(\frac{S}{m_\phi\Lambda^2}\)^{N_c}
=\(\frac{m}{\Lambda}\frac{1+\sqrt{1-4\alpha S}}{2}\)^{N_f}. \label{eqS}
\ee
For $U(2)$ with trace coupling, $W= \half m_\phi\tr\Phi^2+\lambda\tr\Phi$, 
one can easily verify that the result can be obtained by shifting 
\be
m\to m+\lambda/m_\phi, \;\;  {\rm and}\;\;   
\alpha\to \alpha/(1+\frac{\lambda}{mm_\phi})^2.
\ee
Then the equation (\ref{eqS}) with $N_f=N_c$ will produce 
\be
 \quad S=\lambda\Lambda-m_\phi\Lambda^2+mm_\phi\Lambda,
\ee
which is precisely the result in eq.\ (\ref{S}) with $\epsilon=1$.
The $SU(2)$ case can be treated by regarding $\lambda$ as a Lagrange
multiplier and we leave this as an exercise.
The value of the superpotential is 
$N_cS$. 
One should notice that the $S\log S$ term in this case cancelled between 
$W^{s}$ and $W^{d}$.

This confirms, in the simplest possible case, that the method of factorizing  
Seiberg-Witten curve is equivalent to that of minimizing the full superpotential. 
This equivalence is highly nontrivial to directly check  beyond the quadratic superpotential as one can 
see below.

\section{  Superpotential in terms of elliptic integral and Minimization for the cubic}

Here, we calculate the effective superpotential  in terms of elliptic integral. For pure SYM, the calculation is done in \cite{civ}. 
Here we give a treatment with fundamentals for the second simplest case $N_f=2$, $N_c=3$. The values of the physical parameters were already 
determined by the factorization method.  
The superpotential is  
\be
W_{eff}=-\sum_{i=1}^n N_i \Pi_i + \sum_{j=1}^{N_f}D_j
+ 2\pi i (\tau_0 S + \sum_{i=1}^{n-1}b_iS_i) -i(N_c-N_f/2)S +C ,\ee
where
\footnote{Notice the sign difference in $\Pi_i$ integral from 
\cite{csw2} since we perform the integral in first sheet. 
If we change $m\to -m$ as in the usual literature then we need to change the sign of the integral to confirm the known result, as can 
be shown by examining the quadratic potential.}
\be
\Pi_i=\int_{e_i}^{\Lambda} y(z)dz, \;\; 
D_j=\half\int_{m_i}^{\Lambda} y(z)dz, \;\; 
C=\half\((2N_c-N_f)W(\Lambda)+\sum_{I=1}^{N_f} W(z_I)\).  \ee
 
Let 
\be
W'^2+f(x) = g^2(x-e_1)(x-e_2)(x-e_3)(x-e_4), \quad e_1>e_2>e_3>e_4 .\ee 
The integrals are evaluated in terms of elliptic functions; 
\def\sn{{\rm sn}}
\def\dn{{\rm cn}}
\def\cn{{\rm dn}}
\be
S_1=\int_{e_2}^{e_1}y(x)dx 
   =\[2(e_2-e_3)^2\beta\alpha_1^4\]\cdot 
    \int_0^{K_1} \frac{\sn^2u\, \cn^2u\, \dn^2u }
                      {(1-\alpha_1^2\,\sn^2u)^4}
  :=J_1 \cdot I_1, \ee 
\be
S_2=\int_{e_4}^{e_3}y(x)dx 
   =\[2(e_2-e_3)^2\beta\alpha_2^4\]\cdot 
    \int_0^{K_2} \frac{\sn^2u\, \cn^2u\, \dn^2u }
                      {(1-\alpha_2^2\,\sn^2u)^4}
  :=J_2 \cdot I_2, \ee 
\be
\Pi_1= \int_{e_1}^{\Lambda}y(x)dx 
     =\[2(e_1-e_2)^2\beta\alpha_3^4\]\cdot 
      \int_0^{u_\Lambda} \frac{\sn^2u\, \cn^2u\, \dn^2u }
                              {(1-\alpha_3^2\,\sn^2u)^4}
    :=J_3 \cdot I_3, \ee 
\be
\Pi_2= \int_{-\Lambda}^{e_4}y(x)dx 
     =\[2(e_3-e_4)^2\beta\alpha_4^4\]\cdot 
      \int_{u_{(-\Lambda)}}^0 \frac{\sn^2u\, \cn^2u\, \dn^2u }
                                   {(1-\alpha_4^2\,\sn^2u)^4}
    :=J_4 \cdot I_4, \ee 
\be
D_j=\Pi_1(\Lambda)-\Pi_1(m_j), \; for \;\; j=1, ..., N_f,
\ee
where $\beta=\sqrt{(e_1-e_3)(e_2-e_4)}$ and 
\be
\alpha_1^2=\frac{e_1-e_2}{e_1-e_3}<1,\;
\alpha_2^2=\frac{e_3-e_4}{e_2-e_4}<1, \;
\alpha_3^2=\frac{e_1-e_4}{e_2-e_4}>1, \;
\alpha_4^2=\frac{e_1-e_4}{e_1-e_3}>1, \ee and 
$\sn,\cn,\dn$ are Jacobi's elliptic functions \cite{byrd} and 
$u_i$ (=$u$ in $I_i$),  is defined by 
\be
\sn^2u_1=\frac{1}{\alpha_1^2}\frac{x-e_2}{x-e_3},\;
\sn^2u_2=\frac{1}{\alpha_2^2}\frac{e_3-x}{e_2-x},\;
\sn^2u_3=\frac{1}{\alpha_3^2}\frac{x-e_1}{x-e_2},\;
\sn^2u_4=\frac{1}{\alpha_4^2}\frac{e_4-x}{e_3-x}.
\ee
and in each case the $k$'s in the elliptic functions are given by
\be
k_1^2={\alpha_1^2}\frac{e_3-e_4}{e_2-e_4},\;
k_2^2={\alpha_2^2}\frac{e_1-e_2}{e_1-e_3},\;
k_3^2={\alpha_3^2}\frac{e_2-e_3}{e_1-e_3},\;
k_4^2={\alpha_4^2}\frac{e_2-e_3}{e_2-e_4}.
\ee
They satisfy $k_i^2<\alpha_i^2$, $k_i^2<1$. $K_i$ is defined to be $K(k_i)$  
where
\be
K(k)=\frac{\pi}{2}F(\half,\half;1;k^2)
=\frac{\pi}{2}(1+{1\over4}k^2+{9\over64}k^4+...)\ee
satisfies $\sn K_i=1$. $F$ is the hypergeometric function.  
$\sn u_\Lambda=\Lambda$.
The integrals $I_i(\alpha_i)$ look very similar to one another but according 
to whether $\alpha^2$ is bigger or smaller than 1, they have a very different 
behavior as we will see. This corresponds to the difference of $S_i$ and 
$\Pi_i$, the periods along compact vs. non-compact cycles. 
We are interested in the leading orders in a small $\Lambda$ expansion. 
We need to find the $e_i$'s, the zeroes of $y$, in terms of  $m_\phi,m,g,$ 
and the scale of the theory $\Lambda$.
Let $c_1,c_2$ be the classical value of the zeros of $W'$, namely,
$c_{1,2}=-\frac{m_\phi}{2g}\pm\sqrt{\frac{m^2_\phi}{4g^2}-\frac{\lambda}{g}}$. 
\begin{description}
    \item[1] $m_\phi\neq 0$ case:  
    \ba
    e_1&=&c_1+\sqrt{\frac{m-c_1}{c_1-c_2}\epsilon} \Lambda +O(\Lambda^2),\no 
    e_2&=&c_1-\sqrt{\frac{m-c_1}{c_1-c_2}\epsilon} \Lambda +O(\Lambda^2),\no 
e_3&=&c_2+\frac{1}{c_1-c_2}\(4+\frac{m-c_1}{c_1-c_2}\)\epsilon\Lambda^2  
+O(\Lambda^3),\no
e_4&=&a_2=c_2-\frac{2\epsilon\Lambda^2}{c_1-c_2} +O(\Lambda^3).
\ea
If $\lambda=0$,  then $c_1=0, c_2=-m_\phi/g$ and consequently 
\be
e_{1,2}=\pm \sqrt{mg/m_\phi},\;\; e_3=-m_\phi/g 
+g(4+gm/m_\phi)\Lambda^2/m_\phi, \;\; e_4=-m_\phi-2g\Lambda^2/m_\phi.\ee
Here we set $\epsilon=1$.
Now, we can evaluate $J_i$ and $I_i$ in leading order in a small $\Lambda$ 
expansion. 
\be
J_1\sim J_3\sim 8m\Lambda^2, \;\; 
J_2\sim J_4\sim 2g{\Lambda^4\over m_\phi}\(6+gm/m_\phi\)^2, 
I_1 \sim I_2\sim \frac{5\pi}{16}+O(\Lambda^2),\ee
so that $S_i\sim J_i$. 
To obtain $I_3,I_4$, needs more care. 

    \item[2] $m_\phi = 0$ case: 
    In this case,  $c_{1,2}=\sqrt{2\Lambda^2-\lambda/2},$ and especially  if 
$\lambda=0$, $c_{1,2}=\pm\sqrt{2}\Lambda$. 
For the $SU(3)$ case, we have degenerate $c_{1,2}=0$ result. In both cases 
an examination of $W'^2+f=0$ tells us that the leading term is of order 
$m^{1/3}\Lambda^{2/3}$ and the prefactors of $S_i$ and $\Pi_i$ are all 
the same order $J_i \sim m\Lambda^2$ and the integrals $I_i$ for $S_i$ 
are complete elliptic integrals which are of order 1. 
Therefore $W \sim 8m\Lambda^2$.
From this and by dimensional analysis and analyticity of $W$ near $\Lambda=0$, the structure of the leading and subleading orders are 
\be
W \sim m\Lambda^2\( 1+ b_1(\Lambda/m) +b_2(\Lambda/m)^2 +...\) \sim m\Lambda^2 +b_1\Lambda^3+b_2\Lambda^4/m +...
\ee
The first two terms confirm the analysis of the factorization result.  
The main point in this analysis is that  the loop equations say that the first two terms, which are lower order in $\Lambda$,  exist and there is no reason why they should vanish, unlike the one-cut case. They correspond to lower order instanton correction.  It is interesting to see why such terms, forbidden in the unbroken gauge group case can be allowed in the broken cases by examining the gauge theory more closely. 
\end{description}




\begin{thebibliography}{999}


\bibitem{sw}
N.~Seiberg and E.~Witten,
``Monopoles, duality and chiral symmetry breaking in N=2 supersymmetric QCD,''
Nucl.\ Phys.\ B {\bf 431}, 484 (1994)
[arXiv:hep-th/9408099];\\
``Electric - magnetic duality, monopole condensation, and confinement in N=2 
supersymmetric Yang-Mills theory,''
Nucl.\ Phys.\ B {\bf 426}, 19 (1994)
[Erratum-ibid.\ B {\bf 430}, 485 (1994)]
[arXiv:hep-th/9407087].

\bibitem{seiberg}
N.~Seiberg,
``Electric - magnetic duality in supersymmetric nonAbelian gauge theories,''
Nucl.\ Phys.\ B {\bf 435}, 129 (1995)
[arXiv:hep-th/9411149].

\bibitem{kutasov}
D.~Kutasov,
``A Comment on duality in N=1 supersymmetric nonAbelian gauge theories,'' 
Phys.\ Lett.\ B {\bf 351}, 230 (1995)
[arXiv:hep-th/9503086];\\
 D.~Kutasov, A.~Schwimmer, 
``On Duality in Supersymmetric Yang-Mills Theory'', 
Phys.Lett. B354 (1995) 315 
[arXiv:hep-th/9503086];\\
D.~Kutasov, A.~Schwimmer, N. ~Seiberg, 
`Chiral Rings, Singularity Theory and Electric-Magnetic Duality'',
Nucl.Phys. B459 (1996) 455, 
[arXiv:hep-th/9510222].      
\bibitem{kutasovreview}
A.~Giveon and D.~Kutasov,
Rev.\ Mod.\ Phys.\  {\bf 71}, 983 (1999)
[arXiv:hep-th/9802067].


\bibitem{dv}
R.~Dijkgraaf and C.~Vafa,
``A perturbative window into non-perturbative physics,''
[arXiv:hep-th/0208048].

\bibitem{cdsw}
F.~Cachazo, M.~R.~Douglas, N.~Seiberg and E.~Witten,
``Chiral rings and anomalies in supersymmetric gauge theory,''
JHEP {\bf 0212}, 071 (2002)
[arXiv:hep-th/0211170].

\bibitem{dvg} 
R.~Dijkgraaf, M.T. Grisaru, C.S. Lam and C.~Vafa,
``Perturbative calculation of glueball superpotentials,'' 
[arXiv:hepth/0211017].

\bibitem{csw1}
F.~Cachazo, N.~Seiberg and E.~Witten,
``Phases of N = 1 supersymmetric gauge theories and matrices,''
JHEP {\bf 0302}, 042 (2003)
[arXiv:hep-th/0301006].

\bibitem{csw2}
F.~Cachazo, N.~Seiberg and E.~Witten,
``Chiral Rings and Phases of Supersymmetric Gauge Theories,''
JHEP {\bf 0304}, 018 (2003)
[arXiv:hep-th/0303207].


\bibitem{nsw1}
S.~G.~Naculich, H.~J.~Schnitzer and N.~Wyllard,
``The N = 2 U(N) gauge theory prepotential and periods from a  
perturbative matrix model calculation,''
Nucl.\ Phys.\ B {\bf 651}, 106 (2003)
[arXiv:hep-th/0211123].

\bibitem{nsw2}
S.~G.~Naculich, H.~J.~Schnitzer and N.~Wyllard,
``Matrix model approach to the N = 2 U(N) gauge theory with matter 
in the  fundamental representation,''
JHEP {\bf 0301}, 015 (2003)
[arXiv:hep-th/0211254].

\bibitem{civ}
F.~Cachazo, K.~A.~Intriligator and C.~Vafa,
``A large N duality via a geometric transition,''
Nucl.\ Phys.\ B {\bf 603}, 3 (2001)
[arXiv:hep-th/0103067].

\bibitem{cv}
F.~Cachazo and C.~Vafa,
``N = 1 and N = 2 geometry from fluxes,''
[arXiv:hep-th/0206017].

\bibitem{nsw3}
S.~G.~Naculich, H.~J.~Schnitzer and N.~Wyllard,
``Cubic curves from matrix models and generalized Konishi anomalies,''
JHEP 0308, 021 (2003)
[arXiv:hep-th/0303268].

\bibitem{KS}
M.~Klein and S.~J.~Sin,
``On effective superpotentials and Kutasov duality,''
JHEP {\bf 0310}, 050 (2003)
[arXiv:hep-th/0309044].

\bibitem{si}  
N.~Seiberg,
``Exact Results on the Space of Vacua of 
  Four Dimensional SUSY Gauge Theories,''
Phys.\ Rev.\ D {\bf 49}, 685 (1994)
[arXiv:hep-th/9402044];\\
N.~Seiberg, K.~Intriligator,
``Lectures on supersymmetric gauge theories and electric-magnetic duality,''
Nucl.\ Phys.\ Proc.\ Suppl.\ {\bf 45} BC, 1 (1996) 
[arXiv:hep-th/9509066].

\bibitem{murayama} 
C.~Cs\'aki, H.~Murayama,
``Instantons in Partially Broken Gauge Groups,''  
Nucl.\ Phys.\ B {\bf 532}, 498 (1998) 
[arXiv:hep-th/9804061].

\bibitem{janik} Y.~Demasure and R.~A.~Janik,
``Explicit factorization of Seiberg-Witten curves with matter from random matrix models,''
Nucl.\ Phys.\ B {\bf 661}, 153 (2003)
[arXiv:hep-th/0212212].

\bibitem{bala} V.~Balasubramanian, B.~Feng, M.~x.~Huang and A.~Naqvi,
``Phases of N = 1 supersymmetric gauge theories with flavors,''
[arXiv:hep-th/0303065].

\bibitem{Feng} B.~Feng,
``Note on matrix model with massless flavors,''
Phys.\ Rev.\ D {\bf 68}, 025010 (2003)
[arXiv:hep-th/0212274];\\
R.~Roiban, R.~Tatar and J.~Walcher,
``Massless flavor in geometry and matrix models,''
Nucl.\ Phys.\ B {\bf 665}, 211 (2003)
[arXiv:hep-th/0301217].

\bibitem{hananyoz}
A.~Hanany and Y.~Oz,
``On the quantum moduli space of vacua of N=2 supersymmetric $SU(N_c)$ 
gauge theories,''
Nucl.\ Phys.\ B {\bf 452}, 283 (1995)
[arXiv:hep-th/9505075].

\bibitem{bena}
R.~Argurio, V.~L.~Campos, G.~Ferretti and R.~Heise,
``Exact superpotentials for theories with flavors via a matrix integral,''
Phys.\ Rev.\ D {\bf 67}, 065005 (2003)
[arXiv:hep-th/0210291];\\
I.~Bena and R.~Roiban,
``Exact superpotentials in N = 1 theories with flavor and their matrix  model formulation,''
Phys.\ Lett.\ B {\bf 555}, 117 (2003)
[arXiv:hep-th/0211075].

\bibitem{byrd}
Paul F. Byrd and Morris D. Friedman, 
``Handbook of Elliptic integrals for engineers and physicists'', 
Springer-Verlag, 1954.
\end{thebibliography}
\end{document}